\documentclass{optica-article}
\usepackage{eurosym}
\usepackage[colorlinks=true,citecolor=black]{hyperref}

\journal{opticajournal}
\articletype{Research Article}
\begin{document}

\title{Stable multipole solitons in defocusing saturable media with an annular trapping potential}
\author{Xiaoli Lang\authormark{1}, Boris A. Malomed \authormark{2} and Liangwei Dong\authormark{1,*}}
\email{\authormark{*}dlw$\_$0@163.com}

\address{\authormark{1} Department of Physics, Zhejiang University of Science and Technology, Hangzhou,  310023, China \\
	\authormark{2} Instituto de Alta Investigacion, Universidad de Tarapaca, Casilla 7D, Arica, Chile}

\begin{abstract}
We systematically investigate the existence, stability, and propagation dynamics of multipole-mode (necklace-shaped) solitons in the two-dimensional model of an optical medium with the defocusing saturable nonlinearity and an annular potential trough. Various families of stable multipole solitons trapped in the trough, from dipole, quadrupole, and octupole ones to multi-lobe complexes, are found. The existence domain remains invariant with the increase of the potential's depth. Solitons with a large number $N$ of lobes are stable in a wide parameter region, up to $N=48$ and even farther. Actually, stable multipole solitons of an arbitrarily high order $N$ can be found, provided that the trough’s radius is big enough. The power of stable multipoles is essentially larger in comparison to previously studied models. It is demonstrated analytically and numerically that the application of a phase torque initiates stable rotation of the multipole complexes. Thus, we put forward an effective scheme for the stabilization of multipole solitons with an arbitrary high number of lobes, including rotating ones, which offers new possibilities for manipulating complex light beams.
\end{abstract}

%




\section{Introduction}

The stabilization of high-power multipole solitons with a large number of poles (lobes) is a challenging problem in nonlinear optics
\cite{kivshar2003optical,yang20103}, quantum matter (especially Bose-Einstein condensates, BECs), and other physical settings \cite{malomed2022}. The primary difficulty arises from the increasingly strong repulsive interaction between adjacent out-of-phase lobes, which must be balanced by a holding factor.

Multipole (alias \textquotedblleft necklace-shaped” \cite{soljacic1998,soljacic2000}) solitons are characterized by an even number $N$
of lobes (alias \textquotedblleft beads\textquotedblright ) uniformly distributed along a ring. The phase alternation between adjacent lobes
prohibits the existence of odd-$N$ configurations, with the exception of the fundamental one ($N=1$), which corresponds to the axisymmetric ground-state
mode. The simplest structured state is the dipole soliton with $N=2$. While nonlinearity can partly counteract diffraction in the single-component
multipole modes, their long-distance propagation in the free space is ultimately limited by either the radial expansion \cite{soljacic1998} or
spiral instability \cite{soljacic2000}.

Two-dimensional (2D) multipole solitons were first predicted in a two-component system exhibiting focusing saturable nonlinearity~\cite{desyatnikov2001,desyatnikov2002}, where a stable necklace-shaped configuration in one component is maintained by its interaction with a
fundamental mode in the second component, mediated by the XPM (cross-phase-modulation) nonlinearity. On the other hand, the expansion of
necklace beams in focusing Kerr (non-saturable) media can be slowed down by imprinting an angular momentum onto the pattern \cite{soljacic2001}.
Suppression of the expansion of the necklace structures was also demonstrated in systems with fractional diffraction and saturable
nonlinearities \cite{dong2021}. Metastable necklace beams have been predicted in systems with competing quadratic-cubic~\cite{kartashov2002} and
cubic-quintic~(CQ) \cite{mihalache2003,mihalache2004} nonlinearities. Experimental observations of such patterns have been reported in local~\cite{grow2007} and nonlocal~\cite{rotschild2006} nonlinear media.

Photonic lattices (PhLs), induced by spatially-periodic transverse modulations of the local refractive index in optical media, provide a
powerful tool for stabilizing various nonlinear states by means of the effective trapping potential~\cite{yang2004,RevModPhys.83.247,pelinovsky2011localization}. In particular, multipole-mode solitons can be supported by diverse PhL and waveguiding structures~\cite{kartashov20062,yang2004-1,yang20052,rose2007,susanto2008,xia2013,wang20152,
wang201822, kartashov2005, hong20152,kartashov20093}. In addition to multipole solitons supported by spatially periodic \cite%
{yang2004-1,yang20052,rose2007,susanto2008} and quasiperiodic \cite{PhysRevE.74.026601} PhLs, stable nonlinear multipole-modes were
demonstrated in $\mathcal{PT}$-symmetric lattices \cite{wang20152,wang201822}, circular waveguiding arrays \cite{kartashov20093}, and axially symmetric
Bessel PhLs \cite{kartashov2005,dong20092,hong20152}.

Beyond optics, multipole patterns with various values of $N$ have been predicted in BECs featuring both contact~\cite{baizakov2006} and
dipole-dipole~\cite{PhysRevA.97.013636} interactions. In binary BEC systems, the analysis has predicted metastable quantum droplets arranged in
ring-shaped cluster configurations \cite{kartashov20194}. The corresponding system is based on the Gross-Pitaevskii equations with the Lee-Huang-Yang
corrections, that account for beyond-mean-field effects \cite{PhysRevLett.115.155302,PhysRevLett.117.100401}. In particular, stable
multipole quantum droplets can be supported by weakly anharmonic trapping potentials~\cite{dong20223}.

While extensive efforts were put forth to suppress the instability of multipole solitons, arising from the repulsive interaction between adjacent
lobes, the prediction of stable modes with large pole numbers $N$ remains a significant challenge, especially for high-power states. Previous studies
primarily addressed lower-order multipole solitons, particularly dipoles ($N=2$) and quadrupoles ($N=4$). Such stable modes exist in CQ nonlinear
media, with a combination of the 2D harmonic-oscillator trap and a circular Gaussian potential barrier~\cite{liu20232}. However, the respective solitons
with $N\geq 8$ poles exhibit instability in their entire existence domain. Multipole modes with $N=16$ can be made stable in CQ media with an
annular potential, but the respective stability region is very narrow and the corresponding power is very low~\cite{dong2023multipole}.

In this work, we propose a realistic setup that enables the self-trapping of stable solitons with large pole numbers. The combination of the defocusing
saturable nonlinearity and an engineered ring-shaped potential maintains distinct families of multipole solitons, whose radius and radial thickness
can be controlled by adjusting the radius and width of the confining annular potential. The respective stability domain shrinks \emph{very slowly} with
the growth of $N$, permitting stable solitons, at least, up to $N=26$ for typical parameters of the potential. Furthermore, the power of stable
multipole solitons is remarkably enhanced in comparison with the previous studies, which is crucially important for potential applications in
photonics, where one would like to use bright beams. Thus, the most important conclusion is that one can produce stable multipole-mode states
with an arbitrary high order (pole number), $N$, and arbitrarily high power, by selecting a sufficiently large radius of the potential trough. We also
demonstrate robust rotation of the multipole solitons, driven by a phase torque.

The subsequent presentation is structured as follows. The model, based on the 2D Schr\"{o}dinger equation with the self-defocusing saturable
nonlinearity and annular-trough trapping potential, is introduced in Section 2. The results of the systematic analysis are reported in Section 3. The
paper is concluded by Section 4.

\section{The model}

We consider the propagation of optical beams along the $z$-axis in the bulk medium with the saturable defocusing term and a transverse modulation of the
refractive-index profile, which induces an effective axisymmetric trapping potential. The evolution of the optical-wave amplitude $\Psi $ is governed
by the 2D nonlinear Schr\"{o}dinger (NLS) equation, written in the normalized form:
\begin{equation}
\centering i\frac{\partial \Psi }{\partial z}=-\frac{1}{2}\left( \frac{%
\partial ^{2}\Psi }{\partial x^{2}}+\frac{\partial ^{2}\Psi }{\partial y^{2}}%
\right) +\frac{|\Psi |^{2}}{1+s|\Psi |^{2}}\Psi -pV(r)\Psi ,  \label{Eq1}
\end{equation}%
where $s$ quantifies the saturability of the nonlinear response. The transverse $\left( x,y\right) $ and longitudinal $z$ coordinates are scaled,
severally, by the width and diffraction length of the input beam. The trapping annular-trough potential, with $r\equiv \sqrt{x^{2}+y^{2}}$, depth $p>0$, ring radius $r_{0}$, and width $d$, is
\begin{equation}
\centering-pV(r)=-p\exp \left[ -\frac{(r-r_{0})^{2}}{d^{2}}\right] .
\label{Eq2}
\end{equation}%
Equation (\ref{Eq1}) conserves the integral power,
\begin{equation}
U=\iint |\Psi (x,y)|^{2}d\text{x}d\text{y},  \label{U}
\end{equation}%
along with the angular momentum,
\begin{equation}
M=i\iint \Psi ^{\ast }\left( y\frac{\partial \Psi }{\partial x}-x\frac{%
\partial \Psi }{\partial y}\right) d\text{x}d\text{y}  \label{M}
\end{equation}%
($\ast $ stands for the complex conjugate), and Hamiltonian,
\begin{equation}
H=\iint \left\{ \frac{1}{2}\left\vert \nabla \Psi \right\vert ^{2}+\frac{1}{s%
}\left[ \left\vert \Psi \right\vert ^{2}-\frac{1}{s}\ln \left( 1+s\left\vert
\Psi \right\vert ^{2}\right) \right] -pV(r)\left\vert \Psi \right\vert
^{2}\right\} d\text{x}d\text{y}.  \label{H}
\end{equation}

\begin{figure}[tbph]
\par
\begin{center}
\includegraphics[scale=0.5,angle=0]{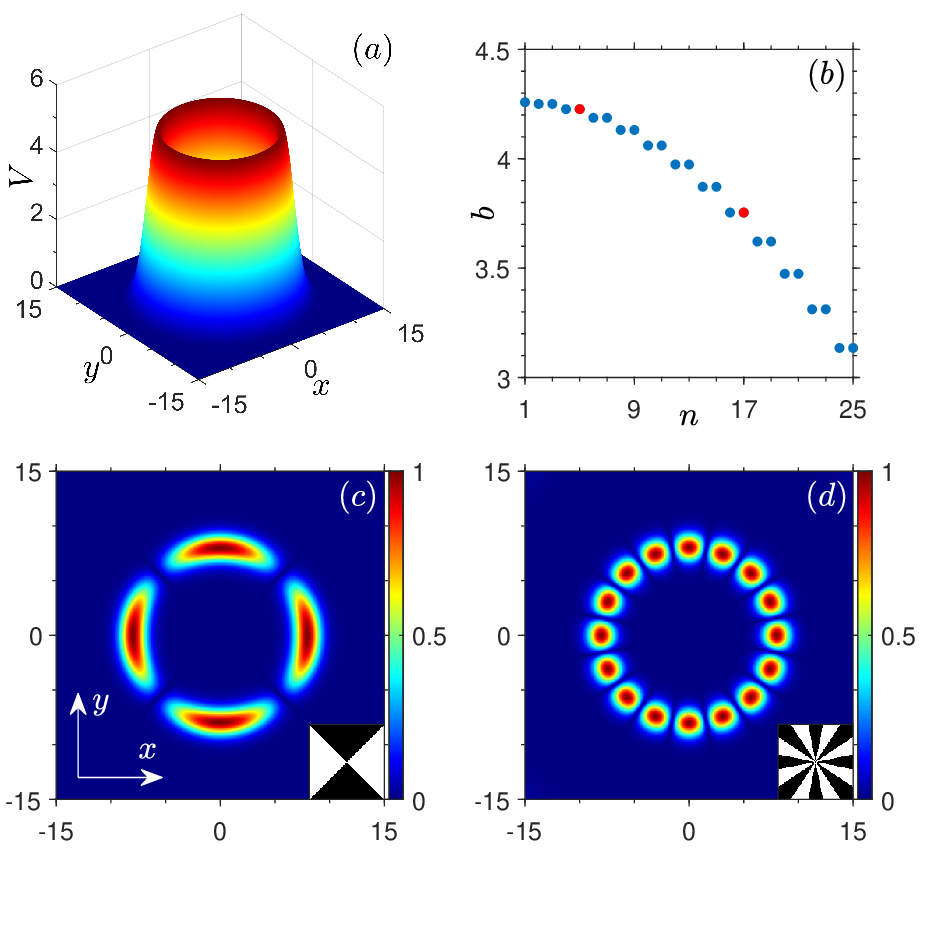} 
\end{center}
\vskip -1.5pc 
\caption{(a) The annular potential, as defined by Eq. (\protect\ref{Eq2}) (it is plotted here with factor $p$). (b) A typical spectrum of discrete
eigenvalues $b$ of the linearized version of equation~(\protect\ref{Eq3}) with this potential. Values of $n$ denote the number of the eigenmode. (c, d)
Absolute values of the wave function, $\left\vert \protect\psi (x,y)\right\vert $, in the quadrupole and $16$-pole linear eigenstates with
eigenvalues marked by two red dots in (b). The insets display the corresponding phase structures. The parameters of the annular potential are $r_{0}=8,d=2$, and $p=5$.}
\label{fig1}
\vskip -1pc 
\end{figure}

Figure~\ref{fig1}(a) shows the ring-shaped potential defined by Eq. (\ref{Eq2}), which is subject to constraint $V(r=r_{0}-r^{\prime
})=V(r=r_{0}+r^{\prime })$ for all $0<r^{\prime }<r_{0}$. This feature plays a crucial role in the stabilization of the nonlinear modes, as it ensures
identical waveguiding conditions in the inner ($r<r_{0}$) and outer ($r>r_{0} $) annular regions. Previous works have demonstrated that, in models
with different nonlinearities, annular potential (\ref{Eq2}) supports three species of stable states: multipole modes \cite{dong2023multipole}, vortex solitons carrying the integer winding number (topological charge) \cite{Dong:23},  and higher-order vortex quantum droplets in binary BECs
\cite{DONG2024114472}. For small $s$, the saturable defocusing nonlinearity in Eq.~(\ref{Eq1}) reduces to the cubic-quintic form with the defocusing cubic ($-|\Psi|^2\Psi$) and focusing quintic ($+s|\Psi|^4\Psi$) terms. This sign combination is fundamentally different from the model in Refs. \cite{dong2023multipole, Dong:23}, which used the focusing cubic ($+|\Psi|^2\Psi$) and defocusing quintic ($-|\Psi|^4\Psi$) terms. Note that the focusing quintic term ($+s|\Psi|^4\Psi$) with small $s$ can be neglected for typical soliton intensities, as its contribution remains small compared to the dominant cubic term.

We search for stationary solutions of Eq.~(\ref{Eq1}) as per the usual ansatz,
\begin{equation}
\Psi (x,y,z)=\psi (x,y)\exp (ibz),  \label{b}
\end{equation}
with real propagation constant $b$ and the transverse field profile, $\psi \left( x,y\right) $. It satisfies the equation
\begin{equation}
\centering\frac{1}{2}\left( \frac{\partial ^{2}\psi }{\partial x^{2}}+\frac{\partial ^{2}\psi }{\partial y^{2}}\right) -b\psi +pV(r)\psi -\frac{|\psi
|^{2}}{1+s|\psi |^{2}}\psi =0,  \label{Eq3}
\end{equation}%
which can be solved by means of iterative methods, \textit{viz}., either the relaxation algorithm or Newton-conjugate-gradient one \cite{yang20103}. The
solutions form continuous families parameterized by propagation constant $b$, saturation parameter $s$, potential depth $p$, and coefficients $r_{0}$ and $d$ of
potential (\ref{Eq2}).

The linear-stability analysis of the stationary solutions can be performed by adding small perturbations, with real and imaginary components, $u(x,y)$
and $v(x,y)$, and complex eigenvalue $\lambda $, to the stationary solution, so that the perturbed solution is
\begin{equation}
\Psi (x,y,z)=\left\{ \psi (x,y)+\left[ u(x,y)+iv(x,y)\right] \exp \left(
\lambda z\right) \right\} \exp (ibz).  \label{pert}
\end{equation}%
The substitution of this in Eq.~(\ref{Eq1}) and linearization with respect to the perturbation yields the eigenvalue problem,
\begin{eqnarray}
&&\lambda u=-\frac{1}{2}\left( \frac{\partial ^{2}}{\partial x^{2}}+\frac{%
\partial ^{2}}{\partial y^{2}}\right) v+bv-pVv+\frac{\psi ^{2}+s\psi ^{4}}{%
(1+s\psi ^{2})^{2}}v  \nonumber \\
&&  \label{Eq4} \\
&&\lambda v=~~\frac{1}{2}\left( \frac{\partial ^{2}}{\partial x^{2}}+\frac{%
\partial ^{2}}{\partial y^{2}}\right) u-bu+pVu-\frac{3\psi ^{2}+s\psi ^{4}}{%
(1+s\psi ^{2})^{2}}u,  \nonumber
\end{eqnarray}%
that can be solved numerically by means of the Fourier collocation algorithm \cite{yang20103}. The underlying soliton is stable if the solution of Eq.~(\ref{Eq4}) yields only imaginary eigenvalues, the instability growth rate (if any) being \textrm{Re(}$\lambda $\textrm{)}.

Prior to producing multipole modes in the framework of the full nonlinear equation (\ref{Eq1}), it is instructive to analyze the spectrum of its
linearized counterpart with the same potential (\ref{Eq2}), as nonlinear modes usually bifurcate from linear eigenstates. The numerically computed
spectrum of the linearized equation comprises a finite set of discrete real eigenvalues, as shown in Fig.~\ref{fig1}(b), along with the obvious
continuous spectrum (not shown here). Note that linear modes with more poles corresponding to the eigenvalues with $n>25$ still exist. Increasing the potential depth $p$ induces a rightward shift in the eigenvalue spectrum, the variation of the thickness $d$ and radius $r_{0}$ also affecting the eigenvalue distribution. The number of guided modes increases with the radius $r_0$. In the limit of $r_0\rightarrow \infty$, the system reduces to the 2D equation with a quasi-1D trapping potential, $V(x)$. While $V(x)$ itself supports only a finite number of trapped transverse modes, the inclusion of a longitudinal plane-wave factor $\exp(iky)$ allows for the construction of infinitely many guided modes, parameterized by the wavenumber $k$.

The spectrum reveals an essential symmetry property: while the ground-state
eigenvalue remains nondegenerate, in agreement with the fundamental principle of quantum mechanics \cite{landau2013quantum}, all excited states
exhibit double degeneracy, with mutually orthogonal eigenmodes sharing identical eigenvalues. The degeneracy implies that the linear eigenmodes
with $N\geq 2$ poles correspond, specifically, to the $N$-th and $N+1$-th eigenvalues. Representative examples of these eigenmodes are displayed in
Figs.~\ref{fig1}(c) and (d) for the fourth and sixteenth excited states, respectively. The linear system permits arbitrary superpositions of the
mutually degenerate eigenmodes as valid solutions. A notable example occurs for the dipole modes ($N=2$), where superpositions of the orthogonal states
generate two vortex solutions with topological charges $m=\pm 1$, demonstrating the direct relationship between the multipole states and those
carrying the angular momentum.

\section{Results and discussion}

\subsection{Dipole and quadrupole solitons}

\begin{figure}[tbph]
\begin{center}
\includegraphics[scale=0.6,angle=0]{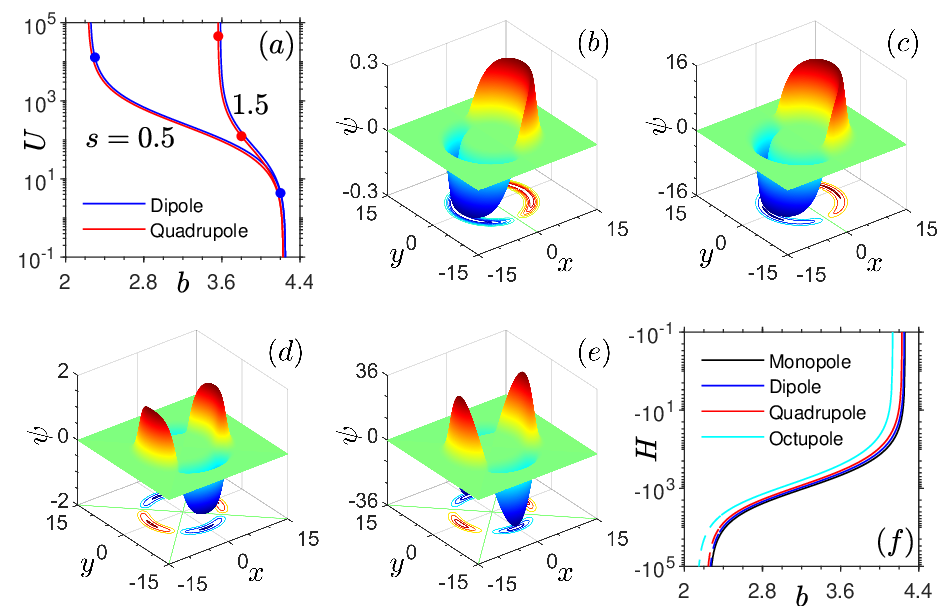}
\end{center}
\vskip -0.5pc 
\caption{(a) Integral power (\protect\ref{U}) versus propagation constant $b$ for dipole and quadrupole solitons in the annular potential (\protect\ref{Eq2}) with the saturation parameters $s=0.5$ and $1.5$. (b, c) Examples of stable low-power ($b=4.2$) and unstable high-power ($b=2.3$) dipole modes marked by blue dots in (a). (d, e) Examples of stable moderate-power ($b=3.8$) and unstable high-power ($b=3.563$) quadrupoles marked by red dots in (a). (f) The Hamiltonian of multipole modes versus $b$. Solid and dashed lines denote the stable and unstable segments. The parameters are $r_{0}=8,d=2$ and $p=5$. $s=0.5$ in (b, c) and $1.5$ in (d, e).}
\label{fig2}
\vskip -0.5pc 
\end{figure}

We now address multipole-mode solitons in the defocusing saturable nonlinear medium. The dipole modes consist of two arc-shaped fragments with the $\pi$-phase difference between them, localized on the ring of radius $r_{0}$, as shown in Figs.~\ref{fig2}(b, c). Both the peak amplitude and thickness of the
dipole solitons monotonously increase with the decrease of the propagation constant $b$. Quadrupole solitons exhibit a more complex structure, composed
of four fragments, with the $\pi $-phase shift between adjacent ones, see Figs.~\ref{fig2}(d, e). For the same potential, the angular span of individual
arc-shaped fragments in the quadrupoles is naturally smaller than in their dipole counterparts.

\begin{figure}[tbph]
\begin{center}
\includegraphics[scale=0.6,angle=0]{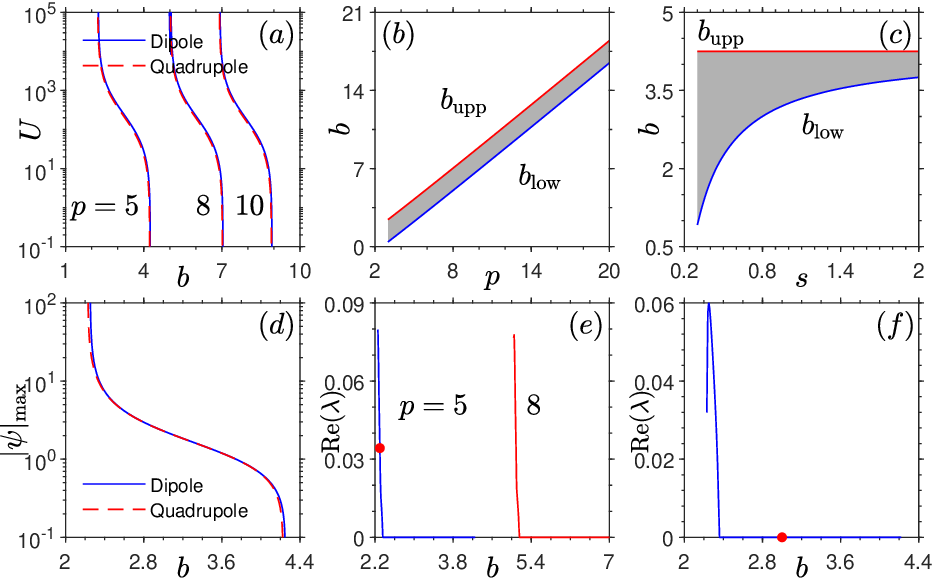} 
\end{center}
 \vskip -0.5pc
\caption{(a) Power $U$ vs. propagation constant $b$ for dipole and quadrupole solitons, for $s=0.5$ and varying values of $p$. (b) The
existence domain of the dipole solitons for $s=0.5$ and varying $p$, which agrees with relation (\protect\ref{s}). (c) The existence domain of the
dipole solitons for $p=5$ and varying values of $s$. (d) The peak value $|\protect\psi |_{\text{max}}$ of the dipole and quadrupole solitons vs. $b$
at $s=0.5,p=5$. (e) The instability growth rate $\text{Re}(\protect\lambda )$ vs. $b$ for the dipole solitons at $s=0.5$. (f) The same for the quadrupole
solitons at $s=0.5,p=5$. Other parameters are $r_{0}=8,d=2$. }
\label{fig3}
 \vskip -0.5pc
\end{figure}

In Fig. \ref{fig2}(a), the soliton families, built according to definition (\ref{b}), stem from the corresponding linear eigenmodes, at upper-cutoff
values of the propagation constant, $b_{\text{upp}}=4.251$ for the dipoles and $b_{\text{upp}}=4.227$ for quadrupoles, which precisely coincide with
the respective linear eigenvalues shown in Fig.~\ref{fig1}(b) (obviously, these values do not depend on the saturation constant $s$). At $s=0.5$, the
power diverges at the lower cutoff, $b_{\text{low}}=2.251$ for the dipoles and $b_{\text{low}}=2.227$ for the quadrupoles. As illustrated below by Fig.~\ref{fig3}(d), the divergence of the integral power is related to diverging values of the local power, $\left\vert \Psi \right\vert ^{2}$, for which the
nonlinear saturable term in Eq. (\ref{Eq1}) is asymptotically replaced by the linear one, $s^{-1}\Psi $, hence the cutoff values of the propagation
constant are obviously related:
\begin{equation}
b_{\text{upp}}=b_{\text{low}}+1/s,  \label{s}
\end{equation}%
which is indeed corroborated by the numerical results [see, e.g., the curves for $s=0.5$ and $1.5$ in Fig.~\ref{fig2}(a)]. This relation is known for
soliton families in other models with saturable nonlinearity, including 1D soliton trains in harmonic PhLs \cite{Kartashov:04} and 2D broken-ring
solitons in Bessel lattices \cite{Dong:08}.

As the potential depth $p$ increases, the existence domain undergoes a systematic rightward shift, see Figs.~\ref{fig3}(a, b), while the shape of
the $U(b)$ dependence remains approximately invariant. This feature is naturally explained by the fact that the eigenvalue of each bound state in a
deep potential well stays at approximately constant distance from the well's bottom. Further, relation (\ref{s}) explains the rapid contraction of the
existence domain for the dipole and quadrupole solitons with the increase of the saturation constant $s$, see Fig.~\ref{fig3}(c).

As seen in Fig.~\ref{fig2}(a), the power of the solitons with different values of $N$ (in that figure, these are $N=2$ and $4$), is a monotonously
decreasing function of propagation constant $b$, hence the families satisfy the \textit{anti-Vakhitov-Kolokolov} (anti-VK) criterion, $dU/db<0$, which
is a necessary stability condition for solitons supported by a defocusing nonlinearity \cite{PhysRevA.81.013624} (the VK criterion per se is opposite,
$dU/db>0$, applying to models with focusing nonlinearity \cite{VK1973, BERGE1998259}). A comprehensive linear-stability analysis of the dipole and
quadrupole soliton families has been performed on the basis of the numerical solution of the eigenvalue problem (\ref{Eq4}), with representative results
plotted in Figs.~\ref{fig3}(e) and (f). For the dipole solitons in the potential with $p=5$, the stable-propagation regime spans $b\in \lbrack
2.35,4.25]$, covering approximately $95\%$ of the total existence domain, $b\in \lbrack 2.25,4.25]$. When the potential depth increases to $p=8$, the
stability region shifts to $b\in \lbrack 5.14,7.04]$ while maintaining the same relative extent ($95\%$) of the respective existence domain, $b\in
\lbrack 5.04,7.04]$ [see Fig.~\ref{fig3}(e)]. Thus, the relative width of the stability region remains approximately invariant with the increase of
the potential depth. Quadrupole solitons exhibit similar stability characteristics, with a stability range of $b\in \lbrack 2.36,4.22]$, that
encompasses about $93\%$ of their existence domain, $b\in \lbrack 2.22,4.22]$ [see Fig.~\ref{fig3}(f)].

\subsection{Multipole solitons}

\begin{figure}[hbtp]
\begin{center}
\includegraphics[scale=0.4,angle=0]{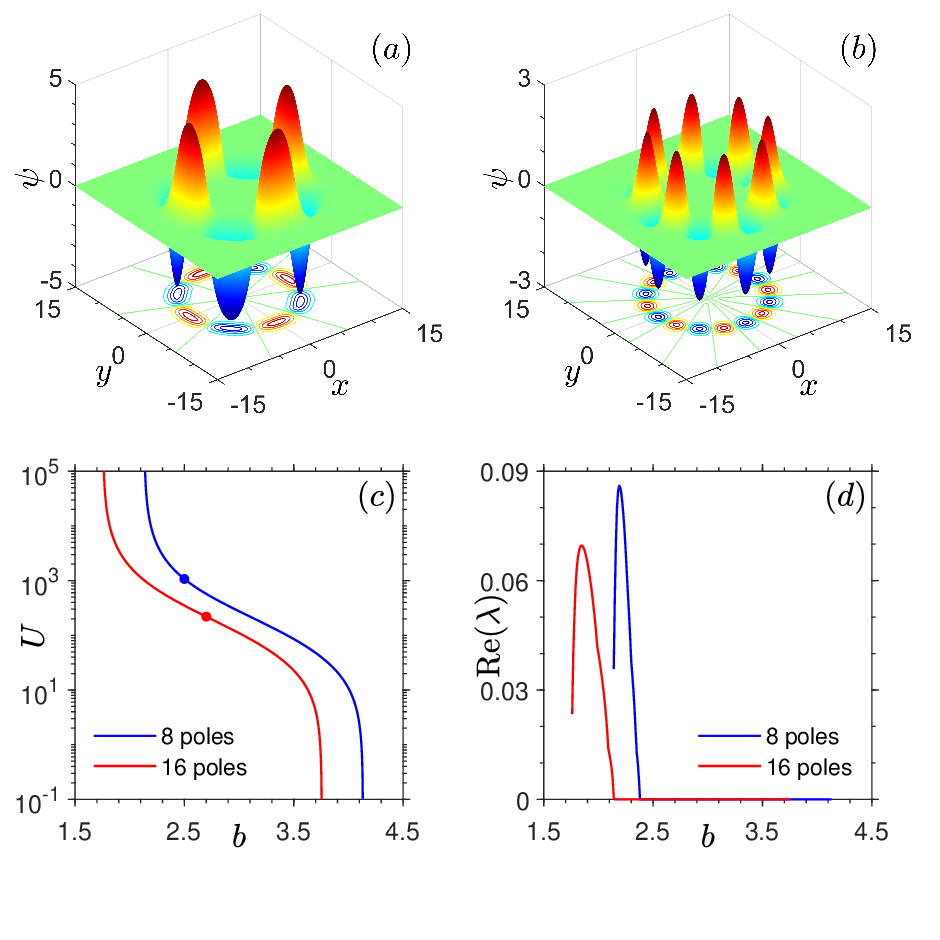} 
\end{center}
\vskip -2pc 
\caption{(a,b) The shapes of 8-pole soliton at $b=2.5$ and 16-pole soliton at $b=2.7$. The corresponding points are marked by dots in panel (c), which
displays integral power $U$ vs. propagation constant $b$ for the 8- and 16-pole solitons. (d) The instability growth rate $\text{Re}(\protect\lambda )$ vs. $b$ for the 8-pole and 16-pole solitons. The parameters are $r_{0}=8,d=2,s=0.5,$ and $p=5$.}
\label{fig4}
\vskip -1pc 
\end{figure}

Next, we consider multipole solitons of a higher order $N$. Figures~\ref{fig4}(a) and (b), respectively, display the stationary field distributions
in the 8-pole and $16$-pole solitons confined within the annular potential trough. Regardless of $N$, all multipole solitons keep the alternating-sign
phase structure between neighboring lobes. Compared to the dipole and quadrupole modes displayed above in Figs.~\ref{fig2}(b-d), these
higher-order solitons naturally exhibit tighter packaging of adjacent fragments. As expected, relation (\ref{s}) holds for them too, see Fig.~\ref{fig4}(c). The large difference between the integral powers of the $8$- and $16$-pole solitons, in comparison to the smaller difference between the
dipoles and quadrupoles [see Fig. \ref{fig2}(a) above] originates from the large separation between the corresponding eigenvalues of the linear
spectrum, as is made evident by Fig.~\ref{fig1}(b).

The key finding of this work is that multipole-mode solitons maintain remarkably broad stability domains across in the underlying parameter space,
even for high pole numbers $N$. In particular, the stability regions of the $8$- and $16$-pole solitons cover, respectively,\ $88\%$ and $80\%$ of their
existence domains, as seen in Fig.~\ref{fig4}(d), in drastic contrast to multipole solitons in the CQ model with the same potential, where the
instability occupies a dominant share of the existence domain \cite{dong2023multipole}. The stability properties reported here are particularly
significant for two reasons: (i) the stability regions for multipole-mode solitons expand steeply; and (ii) there exist stable high-power multipole
solitons, that were previously unattainable.

\begin{figure}[htbp]
	\begin{center}
		\includegraphics[scale=0.4,angle=0]{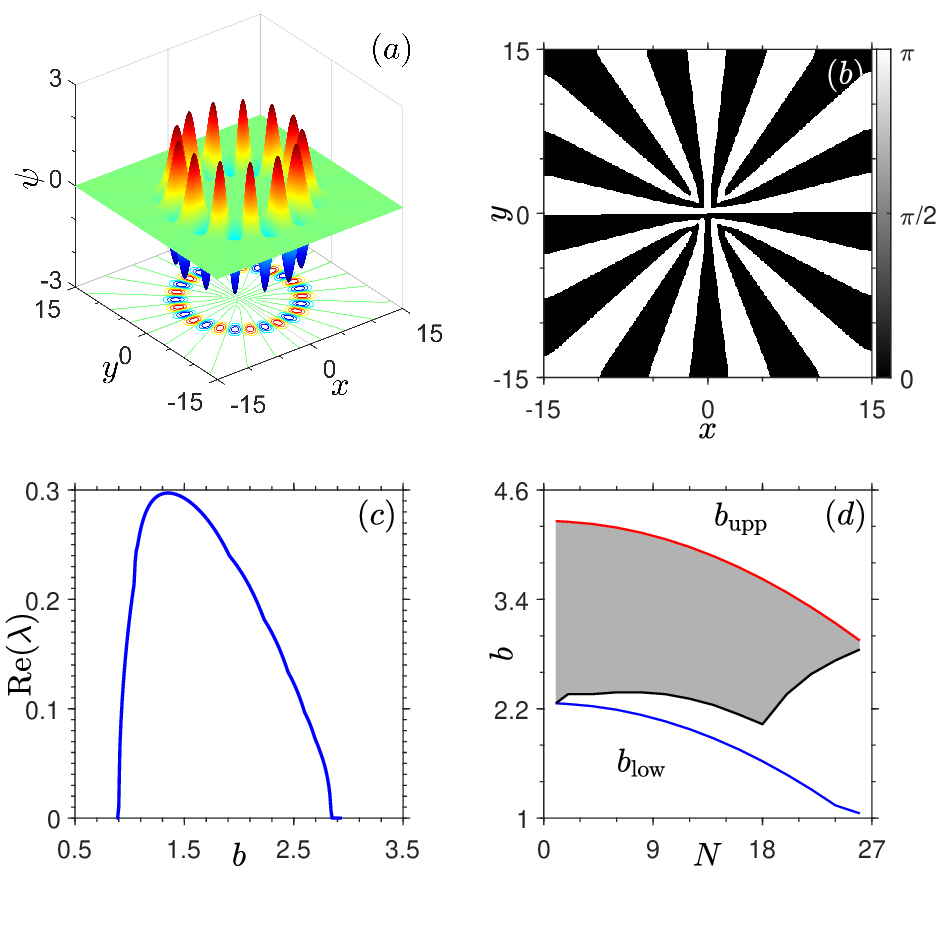}
	\end{center}
	\vskip -1.5pc \vskip -0pc
	\caption{(a, b) The field and phase shapes of the $26$-pole soliton at $b=2$. (c) The instability growth rate $\text{Re}(\protect\lambda )$ vs. $b$ for
		the family of $26$-pole solitons. (d) The existence area (between the $b_{\text{low}}$ and $b_{\text{upp}}$ curves) and the stability region (shaded)
		vs. the pole number $N$. The parameters are $r_{0}=8,d=2,s=0.5$, and $p=5.$}
	\label{fig5}
\end{figure}

To further elucidate properties of the multipole solitons of extremely high orders, we display an example of a\emph{\ }$26$-lobe one in Figs.~\ref{fig5}(a, b). The potential squeezes the lobes along the angular direction, leading to extremely tight packaging and the strong repulsive interaction between
adjacent lobes. Ultimately, it overcomes the trapping capability of the annular potential trough. Therefore, the $26$-pole solitons are unstable in
their nearly entire existence domain, see Fig.~\ref{fig5}(c). Nevertheless, the stability is definitely maintained for the multipoles with $N\leq 18$ in
a substantial part of their parameter space, as shown in Fig.~\ref{fig5}(d). At $N>18$, the stability region quickly shrinks, implying a fundamental
limit for the stability of the higher-order multipole solitons.

\begin{figure}[htbp]
\begin{center}
\includegraphics[scale=0.4,angle=0]{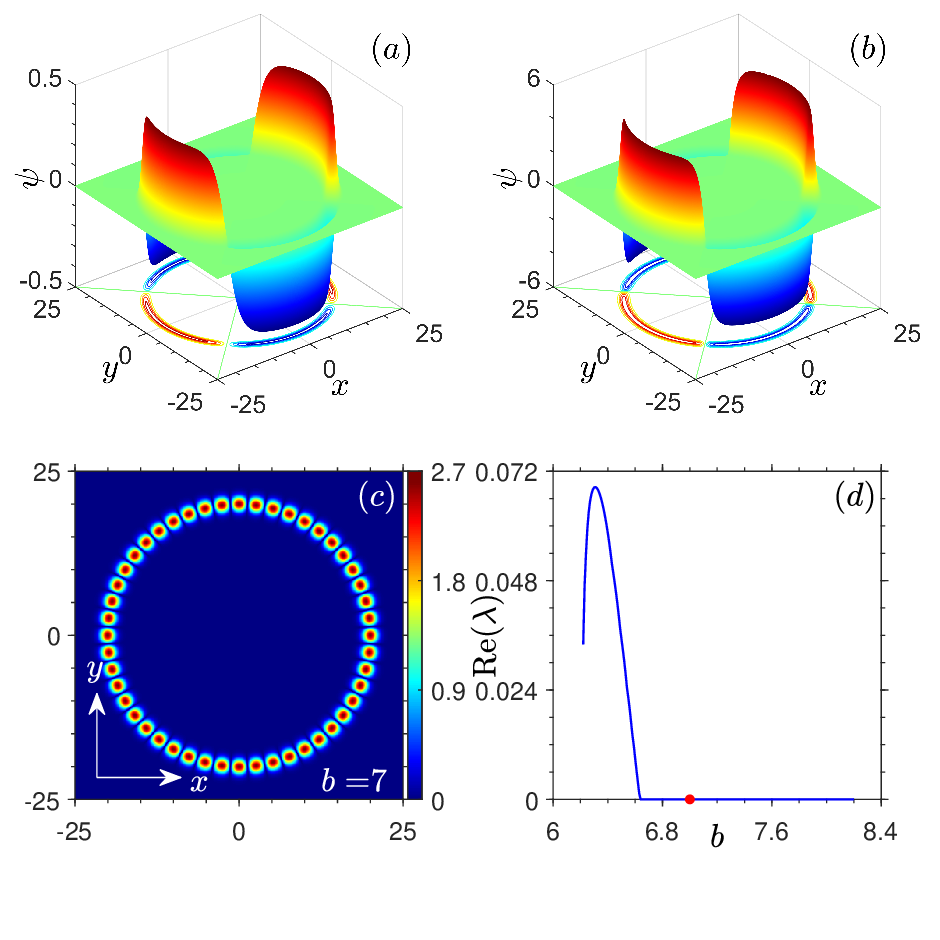}
\end{center}
\par
\vskip -2pc 
\caption{Additional examples of stable multipole solitons trapped in the annular potential trough (\protect\ref{Eq2}) with $r_{0}=20$. (a, b)
Low-power and high-power quadrupole solitons at $b=8.8$ and $7.2$, respectively. (c) The shape of $\left\vert \protect\psi \left( x,y\right)
\right\vert $ in the $48$-pole soliton at $b=7$ marked by the red dot in panel (d), which displays the instability growth rate $\text{Re}(\protect%
\lambda )$ vs. $b$ for the $48$-pole solitons. The parameters are $r_{0}=20,d=2,s=0.5$, and $p=10$.}
\label{fig6}
\end{figure}

As said above, the primary mechanism of the instability of the multipole solitons with a large number $N$ of the lobes originates from the repulsion
between adjacent ones, generating a net outward force that tends to destroy the necklace structure of the circular soliton chain. To counteract this
effect, one can increase radius $r_{0}$ of the annular potential trough (\ref{Eq2}), aiming to reduce the strength of the inter-lobe repulsion, thereby
enabling the generation of stable multipole solitons with still large $N$. To corroborate this possibility, in Fig.~\ref{fig6} we present examples of
stable quadrupole and $48$-pole solitons, which are trapped in the potential with a large radius. In this case, two effects are observed: (i) following
the increase of the integral power, the arc-shaped segments elongate [Figs.~\ref{fig6}(a, b)]; (ii) individual lobes in the necklace-like solitons
exhibit nearly circular profiles, see Fig.~\ref{fig6}(c). Remarkably, in Fig.~\ref{fig6}(d), the $48$-pole solitons remain stable across $79\%$ of
their existence domain, thus representing the highest pole number reported to date for stable multipole solitons. In principle, stable multipole
solitons with arbitrarily high $N$ can be created through the appropriate scaling of the potential-trough radius.

\begin{figure}[tbph]
\begin{center}
\includegraphics[scale=0.4,angle=0]{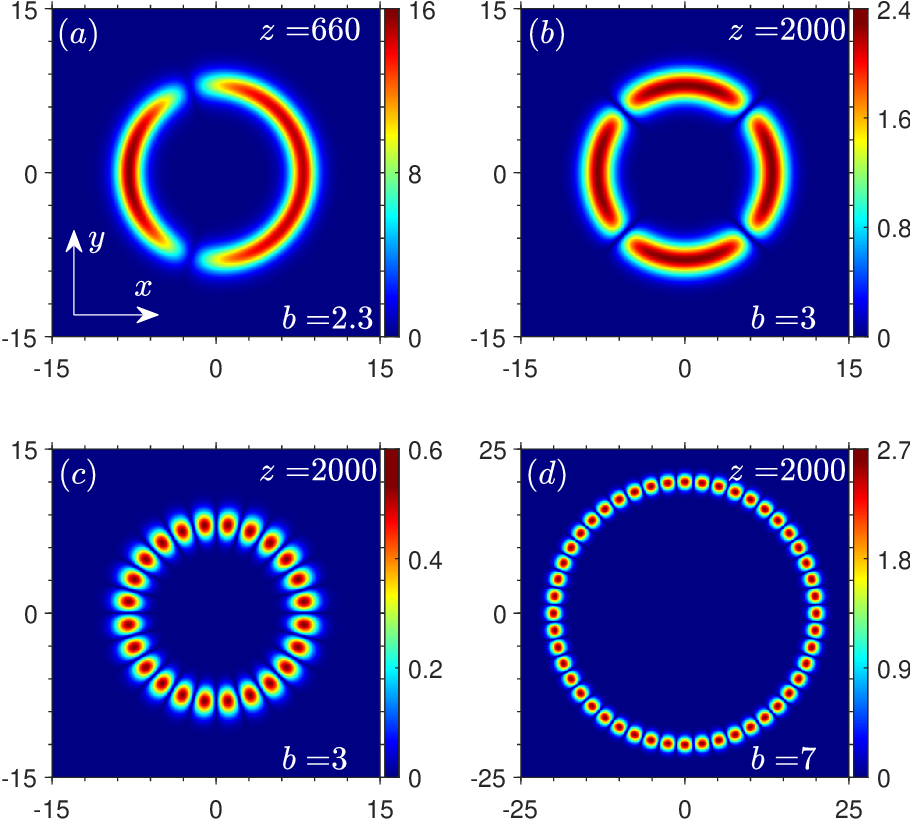} 
\end{center}
\vskip -0.5pc 
\caption{The outcome of the unstable (a) (also see Visualization 1 for the process) and stable (b-d) perturbed propagation of the multipole solitons. The inputs used in (a) and (b) are marked by red dots in Fig.~\protect\ref{fig3}(e) and \protect\ref{fig3}(f), respectively. (c) The outcome of the stable perturbed propagation of the $24$-pole soliton. In panels (a-c) the parameters are $r_{0}=8,d=2,s=0.5$ and $p=5$. (d) The outcome of the stable perturbed propagation of the $48$-pole
soliton marked by the red dot in Fig.~\protect\ref{fig6}(d), with parameters $r_{0}=20,d=2,s=0.5$ and $p=10$.}
\label{fig7}
\vskip -1pc 
\end{figure}

To verify the predictions of linear-stability analysis, we have conducted comprehensive numerical simulations of the perturbed soliton propagation,
using the split-step Fourier method. For the stability test, we added white-noise perturbations to the input at $z=0$. Representative examples of
the evolution are produced in Fig.~\ref{fig7}. At $b=2.3$, the dipole soliton is deformed at $z=660$ [Fig.~\ref{fig7}(a)], which is consistent
with its finite but small instability growth rate [$\mathrm{Re}(\lambda )=0.034$, see Fig.~\ref{fig3}(e)]. The nonzero instability growth rates are complex, indicating the presence of an oscillatory instability. This instability eventually results in oscillations of the two arc-shaped segments on the ring: while one segment shrinks, the other one expands along the ring, swapping their roles in the next cycle of the evolution, see  Visualization 1 for the propagation details. In contrast, the quadrupole and $24$-pole solitons with $\mathrm{Re}(\lambda )=0$ maintain their initial profiles as long as the simulations were running, as shown in Figs.~\ref{fig7}(b, c). The results of the simulation are in full agreement with the predictions of the
linear-stability analysis for all the considered cases, including the solitons with high pole numbers $N$. For instance, the $48$-pole soliton
remains stable at $z=2000$ [Fig.~\ref{fig7}(d)], maintaining its stationary shape intact, cf. Fig.~\ref{fig6}(c). The conclusion is that the perturbed
stable solitons quickly shed off the initially added noise and maintain their structure in the course of indefinitely long evolution. Note that
stable solitons with pole numbers $N\geq 20$ have rarely been found in previously studied models.

\subsection{Rotating multipoles}

\begin{figure}[htbp]
\begin{center}
\includegraphics[scale=0.5,angle=0]{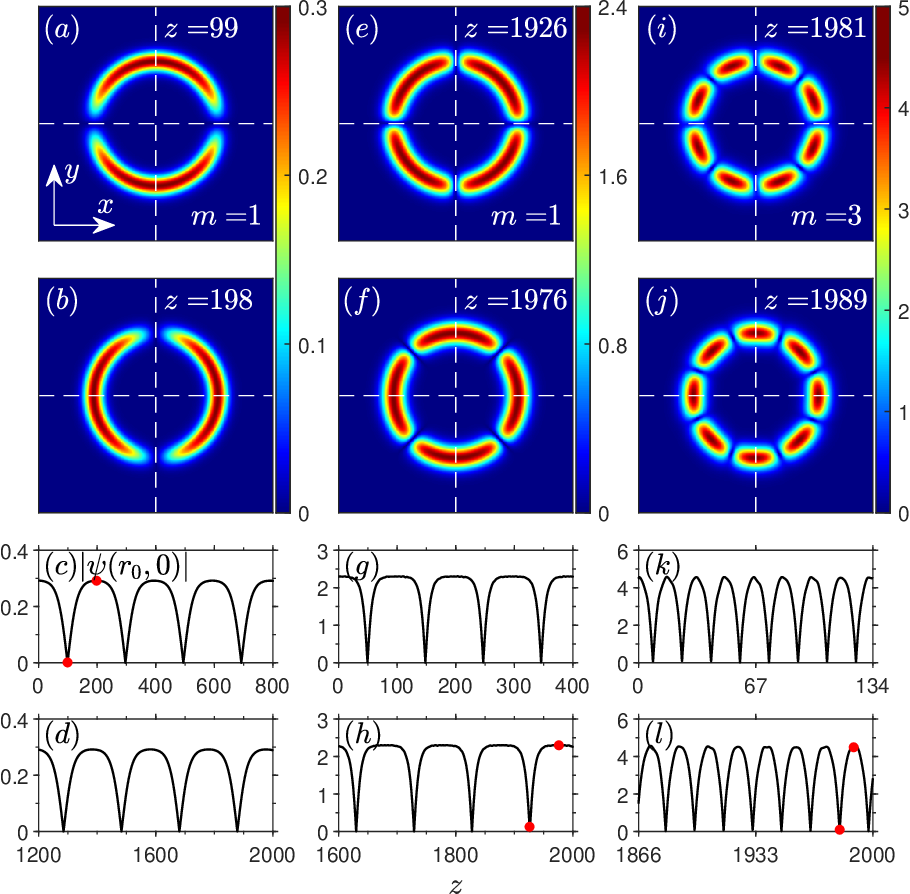}
\end{center}
\vskip -1pc
\caption{The stable counter-clockwise rotation of the dipole with $b=4.2$ (a-d), quadrupole with $b=3$ (e-h), and $8$-pole soliton with $b=2.5$ (i-l),
initiated by the torque kick (\protect\ref{m}), with $m=1$ for dipole and quadrupole, and $m=3$ for the $8$-pole. The initial profiles of the dipole
amd $8$-pole were taken as in Figs.~\protect\ref{fig2}(b) and \protect\ref{fig4}(a), respectively. The bottom panels display the evolution of $%
\left\vert \Psi \left( x,y\right) \right\vert $ at the reference point, $\left( x,y\right) =\left( r_{0},0\right) $, in indicated intervals of the
propagation distance $z$. The configurations displayed in panels (a, b), (e, f) and (i, j) correspond to red dots in (c), (h) and (l), respectively.
The parameters are $r_{0}=8,d=2,s=0.5,$ and $p=5$.}
\label{fig8}
\vskip -1pc
\end{figure}

The well-defined azimuthal structure of the multipole modes considered above makes it possible to set them in rotation by the application of a phase
torque:
\begin{equation}
\psi (x,y)\rightarrow \psi (x,y)\exp (im\theta ),  \label{m}
\end{equation}
where $\theta $ is the angular coordinate, and $m$ is an integer topological charge (winding number). The resulting angular velocity of the rotating
structure can be estimated as $\Omega \approx |m|/r_{0}^{2}$ \cite{dong2023multipole}, the corresponding rotation period being $Z_{\mathrm{est}%
}=2\pi /\Omega \approx 2\pi r_{0}^{2}/|m|$. Numerical simulations confirm this expectation. In particular, the dipole soliton with $r_{0}=8$, kicked
by the torque with $m=1$, exhibits persistent rotation with period $Z_{\mathrm{rot}}\approx 396$ in Figs.~\ref{fig8}(a, b), well matching the
expected value, $Z_{\mathrm{est}}\approx 402$. Further, the simulations for the quadrupole soliton with\ the same parameters, $r_{0}=8$ and $m=1$,
produce the period $Z_{\mathrm{rot}}\approx 400$ in Figs. \ref{fig8}(e, f), which is also very close to $Z_{\mathrm{est}}$. Similarly, the $8$-pole
soliton with $r_{0}=8$ and $m=3$ features $Z_{\mathrm{rot}}\approx 128$ in Figs.~\ref{fig8}(i, j), while the respective estimate is $Z_{\mathrm{est}%
}\approx $ $134$. The rotation observed here, while somewhat similar to the rotating solitons trapped in a multiring potential \cite{YVK2007}, has a fundamentally different origin. In contrast to the whispering-gallery-like modes of \cite{YVK2007}, which are driven by a rotating potential, the multipole rotation here is initiated by the initial torque.  

To demonstrate the robustness of the rotating multipoles, we track the evolution of the respective solution for $\left\vert \Psi \left( x,y\right)
\right\vert $ at a reference point, $(x,y)=(r_{0},0)$. Bottom panels in Fig.~\ref{fig8} represent $5$ full periods of the rotation of the dipole and
quadrupole solitons, initiated by $m=1$ in Eq. (\ref{m}), and $15.6$ periods for the $8$-pole soliton, initiated by $m=3$. In the course of the long
propagation, the soliton profiles fully maintain their structural integrity. The stability of the rotating multispot patterns suggests their potential
application to routing data-carrying optical beams, cf. Ref. \cite{kartashov2005}.

\section{Conclusion}

In summary, we have systematically explored the existence conditions, stability, and propagation dynamics of 2D localized even-numbered multipole
(necklace-shaped) configurations, in the framework of the NLS equation with the defocusing saturable nonlinearity and annular potential trough, in which
the circular configurations are trapped. The multipoles are composed of alternating-phase lobes uniformly distributed along the trough. The
stability domains of the multipole solitons in the parameter space shrink \emph{very slowly} with the growth of the pole number $N$, permitting stable
solitons, at least, up to $N=48$, which is a significant advancement beyond previously reported stability limits for $N$ in other models. In comparison to
the previous studies, the power of the stable multipole solitons is much larger, which is crucially important for potential applications to
photonics. Thus, one can produce \emph{stable} nonlinear multipole states, with an arbitrary large number of poles and arbitrarily high power, using
the annular trapping potential with a sufficiently large radius. The robust rotational dynamics of the multipole solitons, induced by the application of
the phase torque to them, has been demonstrated too.

\vskip 0.5pc
\noindent \textbf{Funding.} National Natural Science Foundation of China (NSFC) (Grant No. 62575264).

\begin{thebibliography}{99}
\bibitem{kivshar2003optical} Y.~S. Kivshar and G.~P. Agrawal, \emph{Optical
solitons: from fibers to photonic crystals} (Academic press, 2003).

\bibitem{yang20103} J.~Yang, \emph{Nonlinear waves in integrable and
nonintegrable systems} (SIAM, 2010).

\bibitem{malomed2022} B.~A. Malomed, \emph{Multidimensional Solitons} (AIP
Publishing LLC, 2022).

\bibitem{soljacic1998} M.~Solja{\v{c}}i{\'c}, S.~Sears, and M.~Segev,
``Self-trapping of ``necklace beams'' in self-focusing {K}err media,'' {%
\JournalTitle{Phys. Rev. Lett.}} \textbf{81}, 4851 (1998).

\bibitem{soljacic2000} M.~Solja{\v{c}}i{\'c} and M.~Segev, ``Self-trapping
of ``necklace-ring'' beams in self-focusing {K}err media,'' {%
\JournalTitle{Phys. Rev. E}} \textbf{62}, 2810 (2000).

\bibitem{desyatnikov2001} A.~S. Desyatnikov, D.~Neshev, E.~A. Ostrovskaya,
Y.~S. Kivshar, W.~Krolikowski, B.~Luther-Davies, J.~J. Garc{\'\i}a-Ripoll,
and V.~M. P{\'e}rez-Garc{\'\i}a, ``Multipole spatial vector solitons,'' {%
\JournalTitle{Opt.
			Lett.}} \textbf{26}, 435--437 (2001).

\bibitem{desyatnikov2002} A.~S. Desyatnikov, D.~Neshev, E.~A. Ostrovskaya,
Y.~S. Kivshar, G.~McCarthy, W.~Krolikowski, and B.~Luther-Davies,
``Multipole composite spatial solitons: theory and experiment,'' {%
\JournalTitle{J. Opt. Soc. Am. B}} \textbf{19}, 586--595 (2002).

\bibitem{soljacic2001} M.~Solja\v{c}i{\'c} and M.~Segev, ``Integer and
fractional angular momentum borne on self-trapped necklace-ring beams,'' {%
\JournalTitle{Phys. Rev. Lett.}} \textbf{86}, 420 (2001).

\bibitem{dong2021} L.~Dong, D.~Liu, W.~Qi, L.~Wang, H.~Zhou, P.~Peng, and
C.~Huang, ``Necklace beams carrying fractional angular momentum in
fractional systems with a saturable nonlinearity,'' {%
\JournalTitle{Commun.
			Nonlinear Sci. Numer. Simul.}} \textbf{99}, 105840 (2021).

\bibitem{kartashov2002} Y.~V. Kartashov, L.-C. Crasovan, D.~Mihalache, and
L.~Torner, ``Robust propagation of two-color soliton clusters supported by
competing nonlinearities,'' {\JournalTitle{Phys. Rev. Lett.}} \textbf{89},
273902 (2002).

\bibitem{mihalache2003} D.~Mihalache, D.~Mazilu, L.-C. Crasovan, B.~A.
Malomed, F.~Lederer, and L.~Torner, ``Robust soliton clusters in media with
competing cubic and quintic nonlinearities,'' {\JournalTitle{Phys. Rev. E}}
\textbf{68}, 046612 (2003).

\bibitem{mihalache2004} D.~Mihalache, D.~Mazilu, L.~C. Crasovan, B.~A.
Malomed, F.~Lederer, and L.~Torner, ``Soliton clusters in three-dimensional
media with competing cubic and quintic nonlinearities,'' {%
\JournalTitle{J. Opt.
			B}} \textbf{6}, S333 (2004).

\bibitem{grow2007} T.~D. Grow, A.~A. Ishaaya, L.~T. Vuong, and A.~L. Gaeta,
``Collapse and stability of necklace beams in {Kerr} media,'' {%
\JournalTitle{Phys.
			Rev. Lett.}} \textbf{99}, 133902 (2007).

\bibitem{rotschild2006} C.~Rotschild, M.~Segev, Z.~Xu, Y.~V. Kartashov,
L.~Torner, and O.~Cohen, ``Two-dimensional multipole solitons in nonlocal
nonlinear media,'' {\JournalTitle{Opt. Lett.}} \textbf{31}, 3312--3314
(2006).

\bibitem{yang2004} J.~Yang, I.~Makasyuk, A.~Bezryadina, and Z.~Chen,
``Dipole and quadrupole solitons in optically induced two-dimensional
photonic lattices: theory and experiment,'' {\JournalTitle{Stud. Appl. Math.}%
} \textbf{113}, 389--412 (2004).

\bibitem{RevModPhys.83.247} Y.~V. Kartashov, B.~A. Malomed, and L.~Torner,
``Solitons in nonlinear lattices,'' {\JournalTitle{Rev. Mod. Phys.}} \textbf{%
83}, 247--305 (2011).

\bibitem{pelinovsky2011localization} D.~E. Pelinovsky, \emph{Localization in
periodic potentials: from Schr{\"{o}}dinger operators to the
Gross--Pitaevskii equation}, (Cambridge University, 2011).

\bibitem{kartashov20062} Y.~V. Kartashov and L.~Torner, ``Multipole-mode
surface solitons,'' {\JournalTitle{Opt. Lett.}} \textbf{31}, 2172--2174
(2006).

\bibitem{yang2004-1} J.~Yang, I.~Makasyuk, A.~Bezryadina, and Z.~Chen,
``Dipole solitons in optically induced two-dimensional photonic lattices,'' {%
	\JournalTitle{Opt. Lett.}} \textbf{29}, 1662--1664 (2004).

\bibitem{yang20052} J.~Yang, I.~Makasyuk, P.~G. Kevrekidis, H.~Martin, B.~A.
Malomed, D.~J. Frantzeskakis, and Z.~Chen, ``Necklacelike solitons in
optically induced photonic lattices,'' {\JournalTitle{Phys. Rev. Lett.}}
\textbf{94}, 113902 (2005).

\bibitem{rose2007} P.~Rose, T.~Richter, B.~Terhalle, J.~Imbrock, F.~Kaiser,
and C.~Denz, ``Discrete and dipole-mode gap solitons in higher-order
nonlinear photonic lattices,'' {\JournalTitle{Appl. Phys. B}} \textbf{89},
521--526 (2007).

\bibitem{susanto2008} H.~Susanto, K.~J.~H. Law, P.~G. Kevrekidis, L.~Tang,
C.~Lou, X.~Wang, and Z.~Chen, ``Dipole and quadrupole solitons in
optically-induced two-dimensional defocusing photonic lattices,'' {%
\JournalTitle{{Physica} D}} \textbf{237}, 3123--3134 (2008).

\bibitem{xia2013} S.~Xia, D.~Song, L.~Tang, C.~Lou, and Y.~Li,
``Self-trapping and oscillation of quadruple beams in high band gap of 2{D}
photonic lattices,'' {\JournalTitle{Chin. Opt. Lett.}} \textbf{11}, 090801
(2013).

\bibitem{wang20152} H.~Wang, S.~Shi, X.~Ren, X.~Zhu, B.~A. Malomed,
D.~Mihalache, and Y.~He, ``Two-dimensional solitons in triangular photonic
lattices with parity-time symmetry,'' {\JournalTitle{Opt. Commun.}} \textbf{%
335}, 146--152 (2015).

\bibitem{wang201822} H.~Wang, X.~Ren, J.~Huang, and Y.~Weng, ``Evolution of
vortex and quadrupole solitons in the complex potentials with saturable
nonlinearity,'' {\JournalTitle{J. Opt.}} \textbf{20}, 125504 (2018).

\bibitem{kartashov2005} Y.~V. Kartashov, R.~Carretero-Gonz\'{a}lez, B.~A.
Malomed, V.~A. Vysloukh, and L.~Torner, \textquotedblleft Multipole-mode
solitons in {B}essel optical lattices,\textquotedblright\ {%
\JournalTitle{Opt. Express}} \textbf{13}, 10703--10710 (2005).

\bibitem{hong20152} W.-P. Hong, ``Surface multipole solitons on
photorefractive media with {B}essel optical lattices,'' {%
\JournalTitle{J.
Korean Phys. Soc.}} \textbf{66}, 919--923 (2015).

\bibitem{kartashov20093} Y.~V. Kartashov, B.~A. Malomed, V.~A. Vysloukh, and
L.~Torner, ``Stabilization of multibeam necklace solitons in circular arrays
with spatially modulated nonlinearity,'' {\JournalTitle{Phys. Rev. A}}
\textbf{80}, 053816 (2009).

\bibitem{PhysRevE.74.026601} H.~Sakaguchi and B.~A. Malomed, ``Gap solitons
in quasiperiodic optical lattices,'' {\JournalTitle{Phys. Rev. E}} \textbf{74%
}, 026601 (2006).

\bibitem{dong20092} L.~Dong, J.~Wang, H.~Wang, and G.~Yin, ``Bessel lattice
solitons in competing cubic-quintic nonlinear media,'' {%
\JournalTitle{Phys. Rev.
			A}} \textbf{79}, 013807 (2009).

\bibitem{baizakov2006} B.~B. Baizakov, B.~A. Malomed, and M.~Salerno,
``Matter-wave solitons in radially periodic potentials,'' {%
\JournalTitle{Phys. Rev. E}} \textbf{74}, 066615 (2006).

\bibitem{PhysRevA.97.013636} C.~Huang, Y.~Ye, S.~Liu, H.~He, W.~Pang, B.~A.
Malomed, and Y.~Li, ``Excited states of two-dimensional solitons supported
by spin-orbit coupling and field-induced dipole-dipole repulsion,'' {%
\JournalTitle{Phys. Rev. A}} \textbf{97}, 013636 (2018).

\bibitem{kartashov20194} Y.~V. Kartashov, B.~A. Malomed, and L.~Torner,
``Metastability of quantum droplet clusters,'' {%
\JournalTitle{Phys. Rev.
Lett.}} \textbf{122}, 193902 (2019).

\bibitem{PhysRevLett.115.155302} D.~S. Petrov, ``Quantum mechanical
stabilization of a collapsing {B}ose-{B}ose mixture,'' {%
\JournalTitle{Phys.
Rev. Lett.}} \textbf{115}, 155302 (2015).

\bibitem{PhysRevLett.117.100401} D.~S. Petrov and G.~E. Astrakharchik,
``Ultradilute low-dimensional liquids,'' {\JournalTitle{Phys. Rev. Lett.}}
\textbf{117}, 100401 (2016).

\bibitem{dong20223} L.~Dong, D.~Liu, Z.~Du, K.~Shi, and W.~Qi, ``Bistable
multipole quantum droplets in binary {B}ose-{E}instein condensates,'' {%
\JournalTitle{Phys. Rev. A}} \textbf{105}, 033321 (2022).

\bibitem{liu20232} D.~Liu, Y.~Gao, D.~Fan, and L.~Zhang, ``Multi-stable
multipole solitons in competing nonlinearity media,'' {%
\JournalTitle{Chaos Soliton.
			Fract.}} \textbf{173}, 113691 (2023).

\bibitem{dong2023multipole} L.~Dong, M.~Fan, C.~Huang, and B.~A. Malomed,
``Multipole solitons in competing nonlinear media with an annular
potential,'' {\JournalTitle{Phys. Rev. A}} \textbf{108}, 063501 (2023).

\bibitem{Dong:23} L.~Dong, M.~Fan, and B.~A. Malomed, ``Stable higher-charge
vortex solitons in the cubic--quintic medium with a ring potential,'' {%
\JournalTitle{Opt. Lett.}} \textbf{48}, 4817--4820 (2023).

\bibitem{DONG2024114472} L.~Dong, M.~Fan, and B.~A. Malomed, ``Stable
higher-order vortex quantum droplets in an annular potential,'' {%
\JournalTitle{Chaos Soliton.
			Fract.}} \textbf{179}, 114472 (2024).

\bibitem{landau2013quantum} L.~D. Landau and E.~M. Lifshitz, \emph{Quantum
mechanics: non-relativistic theory}, vol.~3 (Elsevier, 2013).

\bibitem{Kartashov:04} Y.~V. Kartashov, V.~A. Vysloukh, and L.~Torner,
``Soliton trains in photonic lattices,'' {\JournalTitle{Opt. Express}}
\textbf{12}, 2831--2837 (2004).

\bibitem{Dong:08} L.~Dong, J.~Wang, H.~Wang, and G.~Yin, ``Broken ring
solitons in {B}essel optical lattices,'' {\JournalTitle{Opt. Lett.}} \textbf{%
33}, 2989--2991 (2008).

\bibitem{PhysRevA.81.013624} H.~Sakaguchi and B.~A. Malomed, ``Solitons in
combined linear and nonlinear lattice potentials,'' {%
\JournalTitle{Phys.
Rev. A}} \textbf{81}, 013624 (2010).

\bibitem{VK1973} N.~G. Vakhitov and A.~A. Kolokolov, ``Stationary solutions
of the wave equation in the medium with nonlinearity saturation,'' {%
\JournalTitle{Radiophys. Quant. El+}} \textbf{16}, 783--789 (1973).

\bibitem{BERGE1998259} L.~Berg\'e, ``Wave collapse in physics: principles
and applications to light and plasma waves,'' {\JournalTitle{Phys. Rep.}}
\textbf{303}, 259--370 (1998).

\bibitem{YVK2007}  Y.~V. Kartashov,V.~A. Vysloukh, and L.~Torner,, ``Rotating surface solitons,'' {%
	\JournalTitle{Opt. Lett.}} \textbf{32}, 2948--2950 (2007).

\end{thebibliography}
\vskip 0.5pc
\noindent  \textbf{Disclosures.}  The authors declare no conflicts of interest.
\vskip 0.5pc
\noindent \textbf{Data availability.}  Data underlying the results presented in this paper are not publicly available at this time but may
be obtained from the authors upon reasonable request.

\end{document}